\documentclass[paper,twoside,notoc]{JHEP}
\usepackage{epsfig}
\def\Im{\mathop{\mbox{Im}}}
\def\Re{\mathop{\mbox{Re}}}

\newcommand{\bea}{\begin{eqnarray}}
\newcommand{\be}{\begin{equation}}
\newcommand{\eea}{\end{eqnarray}}
\newcommand{\ee}{\end{equation}}
\newcommand{\nn}{\nonumber}
\newcommand{\db}{\bar{d}}
\newcommand{\gd}{\gamma_\mu}
\newcommand{\gu}{\gamma^\mu}

\newcommand{\als}{\alpha_s}

\preprint{TUM-HEP-320/98\\
  ROME1-1217/98\\
  Edinburgh 98/14\\
  ROM2F/98/33} 
\title{$\Delta M_K$ and $\varepsilon_K$ in SUSY at the
  Next-to-Leading order} 
\author{M. Ciuchini and V. Lubicz\\Dip. di Fisica, Universit\`{a} 
  di Roma Tre and\\
  INFN, Sezione di Roma III, Via della Vasca Navale 84, I-00146 Roma,
  Italy} 
\author{L. Conti and A. Vladikas\\INFN, Sezione di Roma II, and
  Dip. di Fisica, Univ. di Roma ``Tor Vergata''\\
  Via della Ricerca Scientifica 1, I-00133 Roma, Italy}
\author{A. Donini\\Dep. de Fisica Teorica, Univ. Autonoma Madrid,\\
  Fac. de Ciencias, C-XI, Cantoblanco, E-28049 Madrid, Spain}
\author{E. Franco, G. Martinelli and I. Scimemi\\Dipartimento di
  Fisica, Universit\`a di Roma ``La
  Sapienza''\\and INFN, Sezione di Roma, P.le A. Moro, I-00185 Roma,
  Italy}
\author{V. Gimenez\\Dep. de Fisica Teorica and IFIC, Univ. de Valencia,\\
  Dr. Moliner 50, E-46100, Burjassot, Valencia, Spain}
\author{L. Giusti\\Scuola Normale Superiore, P.zza dei Cavalieri 7 and\\
  INFN, Sezione di Pisa, 56100 Pisa, Italy}
\author{A. Masiero\\SISSA-ISAS, Via Beirut 2-4, I-34013 Trieste, Italy and\\
  INFN, Sezione di Trieste, Trieste, Italy} 
\author{L. Silvestrini\\Physik Department, Technische Universit\"{a}t
  M\"{u}nchen,\\D-85748 Garching, Germany} 
\author{M. Talevi\\Department of Physics \& Astronomy, University of
  Edinburgh,\\The King's Buildings, Edinburgh EH9 3JZ, UK} 

\abstract{We perform a Next-to-Leading order analysis of $\Delta S=2$
  processes beyond the Standard Model. Combining the recently computed
  NLO anomalous dimensions and the $B$ parameters of the most general
  $\Delta S=2$ effective Hamiltonian, we give an analytic formula for
  $\Delta M_K$ and $\varepsilon_K$ in terms of the Wilson coefficients
  at the high energy scale. This expression can be used for any
  extension of the Standard Model with new heavy particles. Using this
  result, we consider gluino-mediated contributions to $\Delta S=2$
  transitions in general SUSY models and provide an improved analysis
  of the constraints on off-diagonal mass terms between the first two
  generations of down-type squarks. Finally, we improve the constraints
  on R-violating couplings from $\Delta M_K$ and $\varepsilon_K$.}


\begin{document}

\section{Introduction}
\label{sec:intro}  

The prescription of minimality in the number of new particles
introduced to supersymmetrise the Standard Model (SM), together with
the demand of conservation of baryon and lepton numbers, does not
prevent the appearance of more than 100 new SUSY parameters in
addition to the 18 already present in the SM. Fortunately for SUSY
predictivity, most of this enormous parameter space is already
phenomenologically ruled out. Flavour changing neutral current (FCNC)
and CP violating phenomena are the protagonists of this drastic
reduction of the SUSY degrees of freedom. Therefore, the interest on
the FCNC constraints on SUSY is well motivated: given the large degree
of arbitrariness in the construction of low-energy SUSY models, they
represent one of the main criteria we have for selecting viable
theories. At the same time, they can shed some, albeit dim, light on
the more fundamental physics from which low-energy SUSY stems.

The ignorance of the underlying physics compels us to find ways to
analyse the impact of FCNC and CP violating processes on low-energy
SUSY in a model-independent manner. Since the pioneering work of Hall,
Kostelecky and Raby in 1986 \cite{hall}, the most adequate tool for
such analyses has proven to be the so-called mass insertion
approximation. A brief description of this method will be provided
later on. We note here that this method has been applied, so far,
mainly to the most genuine SUSY sources of FCNC, i.e. the Flavour
Changing (FC) couplings of gluinos and neutralinos to fermions and
sfermions \cite{duncan}.  Differently from the other sources of FCNC
in SUSY (namely W, charged Higgs and chargino exchanges), these
couplings have no analogue in the SM.

As long as we deal with general squark mass matrices, the inclusion of
gluino-mediated FCNC diagrams alone is sufficient to get the correct
bulk of the SUSY contribution. In specific models, corresponding to
particularly restricted squark masses, it may occur, however, that
other contributions become more important. This is the case, for example,
of the Constrained Minimal Supersymmetric Standard Model (CMSSM) with
exact universality of the soft breaking terms: in this case, in
regions of the SUSY parameter space where charginos, stops and/or
charged Higgs are relatively light, their exchange may overwhelm the
gluino contribution to FCNC (see for instance \cite{pokorski}). In
this work we intend to study the general case of squark mass matrices
and, thus, we concentrate on the gluino and squark exchange
contributions. We postpone the inclusion of the chargino-squark
contributions in the mass-insertion framework to a subsequent work.

In the literature there exist four analyses which make use of the
abovementioned method to perform a complete study of the whole variety
of FCNC processes in the hadronic and leptonic sectors
\cite{pokorski}--\cite{GGMS}. These studies do not include QCD
radiative corrections and make use of the Vacuum Insertion
Approximation (VIA) in the evaluation of the hadronic matrix elements.
A major step forward was recently made by the inclusion of the leading
QCD corrections in the evaluation of the gluino-exchange contributions
to $K^0$--$ \bar K^0$ mixing \cite{bagger}. What Bagger et al. found
is that the leading QCD corrections affect the results in a
non-negligible way: in particular for comparable values of the squark
and gluino masses, the QCD corrections increase the lower bound on the
squark masses of the first two generations by a factor
three.  As for the $B$-parameters of the hadronic matrix elements, the
VIA values were used.

Motivated by the encouraging results of ref.~\cite{bagger}, by the
recent computation of the Next-to-Leading (NLO) QCD corrections to the
most general $\Delta F = 2$ effective Hamiltonian ${\cal H}_{\rm
  eff}^{\Delta F = 2}$ \cite{ciuchini} and by the lattice calculation
of the full set of $B$-parameters contributing to the $K^0$ mixing
matrix elements \cite{roma, otherb}, we present here the results of a new
analysis of $K^0$--$\bar K^0$ mixing which includes these progresses
intended to improve the theoretical accuracy. 

A full NLO computation needs the evaluation of the $O(\alpha_s)$
corrections to the Wilson coefficients at the scale of the SUSY masses
running in the loop. This piece of information is not available yet
for gluino contributions\footnote{NLO QCD corrections to the matching
  conditions for charged-Higgs and chargino contributions to
  $\varepsilon_K$ have been very recently computed \cite{krauss}.}.
We can only invoke the smallness of $\alpha_s$ evaluated at such large
scale as a (arguable) hint for the relative smallness of these
corrections. Even with this limitation, we think that it is relevant
to provide the improvement that the knowledge of the NLO QCD
corrections to ${\cal H}_{\rm eff}^{\Delta S =2}$ and the lattice
values of the $B$-parameters allow us to obtain.

A final comment is in order. One may wonder whether such a refinement
(inclusion of NLO QCD corrections and $B$-parameters from lattice
computations) is worth in a study where many uncertainties, due to the
presence of undetermined SUSY parameters and to the use of certain
approximations (in particular the omission of the full set of FCNC
contributions with possible non-negligible interference effects), are
present. We think that the answer is positive at least for two
reasons. The practical one is that, due to the complete
model-independence of our approach, it is possible to ``test'' rapidly
the impact of FCNC processes on new classes of low-energy SUSY models
by just comparing their predictions for the FC parameters $\delta$
(see below) with the constraints that we provide. Obviously a
determination as precise as possible of these $\delta$s is welcome.
The second motivation is that with the future inclusion of the
chargino exchange contributions to FCNC processes we will be able to
obtain a comprehensive framework where to perform general studies of
low-energy SUSY. In view of this, the use of ${\cal H}_{\rm
  eff}^{\Delta F=2}$
consistently renormalised at the NLO and a better determination of the
$B$-parameters play a major role in obtaining an efficient tool of
analysis. Our aim is to approach as much as possible the situation of
the computation of $\Delta F =2$ processes in the SM, where the
complete LO \cite{gilman} and NLO \cite{buras} analyses have been
carried out. 

To this aim, we provide an analytic formula for the NLO ${\cal H}_{\rm
  eff}^{\Delta S=2}$ in terms of the Wilson coefficients at the SUSY
scale. This expression, together with the values of the
$B$-parameters, can be readily used to compute $\Delta M_K$ and
$\varepsilon_K$ in any given SUSY model. Indeed, the formula we give
is valid in any extension of the SM in which new heavy particles
contributing to FCNC processes are present, since the basis of
operators considered here is the most general for $\Delta S =2$
transitions. One just has to plug in the initial conditions computed
in his favourite model. At the moment, the full NLO expression for
${\cal H}_{\rm eff}^{\Delta S=2}$, including the $O(\alpha_s)$
contributions to the matching conditions, is only available for the SM
\cite{buras}. The contributions to $\varepsilon_K$ (i.e. to the
imaginary part of the Hamiltonian) are also known for the two Higgs
doublets model \cite{soff} and for the chargino contribution in the
CMSSM \cite{krauss}.  In all these cases, the $O(\alpha_s)$ terms in
the matching amount to a very small correction, and the bulk of the
effect is due to the running down to the hadronic scale. In our case,
we find that the effect of NLO corrections to the running of the
Left-Right operators is quite large (of the order of 30\% or more),
which is somewhat worrying with respect to the convergence of the
renormalisation-group improved perturbative expansion.

The paper is organised as follows. Next section will introduce the
${\cal H}_{\rm eff}^{\Delta S=2}$ in SUSY with the inclusion of the
NLO QCD corrections in the evolution from the SUSY scale down to low
energy. We will also provide the expression of the Wilson
coefficients at the hadronic scale (to be detailed below) as a
function of the Wilson coefficients and of $\alpha_s$ at the SUSY
scale.  Section \ref{sec:ME} deals with the evaluation of the hadronic
matrix elements of the local operators of ${\cal H}_{\rm
  eff}^{\Delta S=2}$. Here we will replace the ``traditional'' values
of the $B$-parameters in the VIA with the values which were recently
obtained in a lattice computation \cite{roma}. Our quantitative
results are presented in Sect.~\ref{sec:numerics} and summarised in
the Tables \ref{tab:reds2_200}--\ref{tab:imds2_1000}. These latter
contain new constraints on the FC parameters $\delta$ and a
comparison with the previous results. The major effect of our improved
computation is felt by Left-Right operators. As a further
application, in Sect.~\ref{sec:R}, we consider the contribution to 
$K^0 $--$\bar K^0$ mixing in models with explicit R parity and lepton
number violations. We obtain new bounds for the relevant R parity
breaking $\lambda'$-type couplings and compare them to those which
were previously obtained in the VIA case and without QCD corrections.
Finally Sect.~\ref{sec:concl} contains some conclusions and an
outlook.

\section{Effective Hamiltonian for $\Delta S=2$ processes in SUSY}
\label{sec:EH}

In this Section, we describe the computation of gluino-mediated
contributions to ${\cal H}_{\rm eff}^{\Delta S=2}$ at the NLO in QCD.

In ref.~\cite{bagger}, ${\cal H}_{\rm eff}^{\Delta S=2}$ was computed
at the LO in three different cases: i) $m_{\tilde{q}} \sim
m_{\tilde{g}}$, ii) $m_{\tilde{q}} \ll m_{\tilde{g}}$ and iii)
$m_{\tilde{q}} \gg m_{\tilde{g}}$ , where $m_{\tilde{q}}$ is the
average squark mass and $m_{\tilde{g}}$ is the gluino mass.

Case ii), in which $m_{\tilde{q}} \ll m_{\tilde{g}}$, cannot be
realized in the framework considered here, in which the soft SUSY
breaking terms are introduced at the Planck scale. This is due to the
fact that the evolution from the Planck to the electroweak scale
forbids such a mass hierarchy. In fact, neglecting the effects of
Yukawa couplings and A-terms, one obtains in the down sector the
following approximate expression for the ratio
$x=m^2_{\tilde{g}}/m_{\tilde{q}}^2$ at the electroweak scale, in terms
of the value $x_0$ at the Planck scale \cite{louis}:
\begin{equation}
  \label{xlimit}
  x\simeq \frac{9 x_0}{1+7 x_0} \longrightarrow \frac{9}{7}\, .
\end{equation}
Even if one starts at the superlarge scale with an extreme hierarchy
between squark and gluino masses ($x_0 \gg 1$), at the electroweak
scale the two masses will be of the same order.  For this reason, we
will not consider the case $m_{\tilde{q}} \ll m_{\tilde{g}}$ in our
analysis.  Case iii), in which $m_{\tilde{q}} \gg m_{\tilde{g}}$, can
be realized in some special models, such as effective supersymmetry
\cite{kaplan} or models with a light gluino \cite{farrar}. However,
due to the peculiar features and signatures of these models, a careful
analysis of these cases would lie beyond the scope of this paper (see
ref.~\cite{ignazio} for a NLO analysis of $\Delta S=2$ processes in
these models).  Therefore, in the following we will consider the case
$m_{\tilde{q}} \sim m_{\tilde{g}}$, and compare our results with the
zeroth- \cite{GGMS} and leading-order \cite{bagger} results previously
published.

We will perform our computation in the so-called mass insertion
approximation \cite{hall}.  One chooses the super-CKM basis for the
fermion and sfermion states, where all the couplings of these particles
to neutral gauginos are flavour diagonal, while the genuine SUSY FC
effect is exhibited by the non-diagonality of the sfermion mass
matrices. Denoting by $\Delta^2$ the off-diagonal terms in the sfermion
mass matrices (i.e.  the mass terms relating sfermions of the same
electric charge, but different flavour), the sfermion propagators can
be expanded as a series in terms of $\delta = \Delta^2/ \tilde{m}^2$,
where $\tilde{m}$ is the average sfermion mass.  As long as $\Delta^2$
is significantly smaller than $\tilde{m}^2$, we can just take the
first term of this expansion and, then, the experimental information
concerning FCNC and CP violating phenomena translates into upper
bounds on the $\delta$s \cite{gabbiani}--\cite{GGMS}. 

Obviously the above mass insertion method presents the major advantage
that one does not need the full diagonalisation of the sfermion mass
matrices to perform a test of the SUSY model under consideration in
the FCNC sector. It is enough to compute ratios of the off-diagonal
over the diagonal entries of the sfermion mass matrices and compare
the results with the general bounds on the $\delta$s which we provide
here from $\Delta M_K$ and $\varepsilon_K$. This formulation of the
mass insertion approximation in terms of the parameters $\delta$ is
particularly suitable for model-independent analyses, but involves two
further assumptions: the smallness of the off-diagonal mass terms with
respect to the diagonal ones, and the degeneracy of the diagonal mass
terms in the super-CKM basis. The latter assumption, related to the
use of the average squark mass $\tilde{m}$, is well justified in our
case, since, on quite general grounds, one does not expect a sizeable
non-degeneracy of the first two generations of down-type squarks.
It is also possible, however, to define the mass insertion
approximation in a more general way, which is also valid for
non-degenerate diagonal mass terms (see ref.~\cite{romanino} for
details).

There exist four different $\Delta$ mass-insertions connecting
flavours $d$ and $s$ along a sfermion propagator:
$\left(\Delta^d_{12}\right)_{LL}$, $\left(\Delta^d_{12}\right)_{RR}$,
$\left(\Delta^d_{12}\right)_{LR}$ and
$\left(\Delta^d_{12}\right)_{RL}$.  The indices $L$ and $R$ refer to
the helicity of the fermion partners.  The amplitude for $\Delta S=2$
transitions in the full theory at the LO is given by the computation
of the diagrams in fig.~\ref{fig:ds2}.

\FIGURE{     
\epsfysize=12cm 
\epsfxsize=14cm 
\epsffile{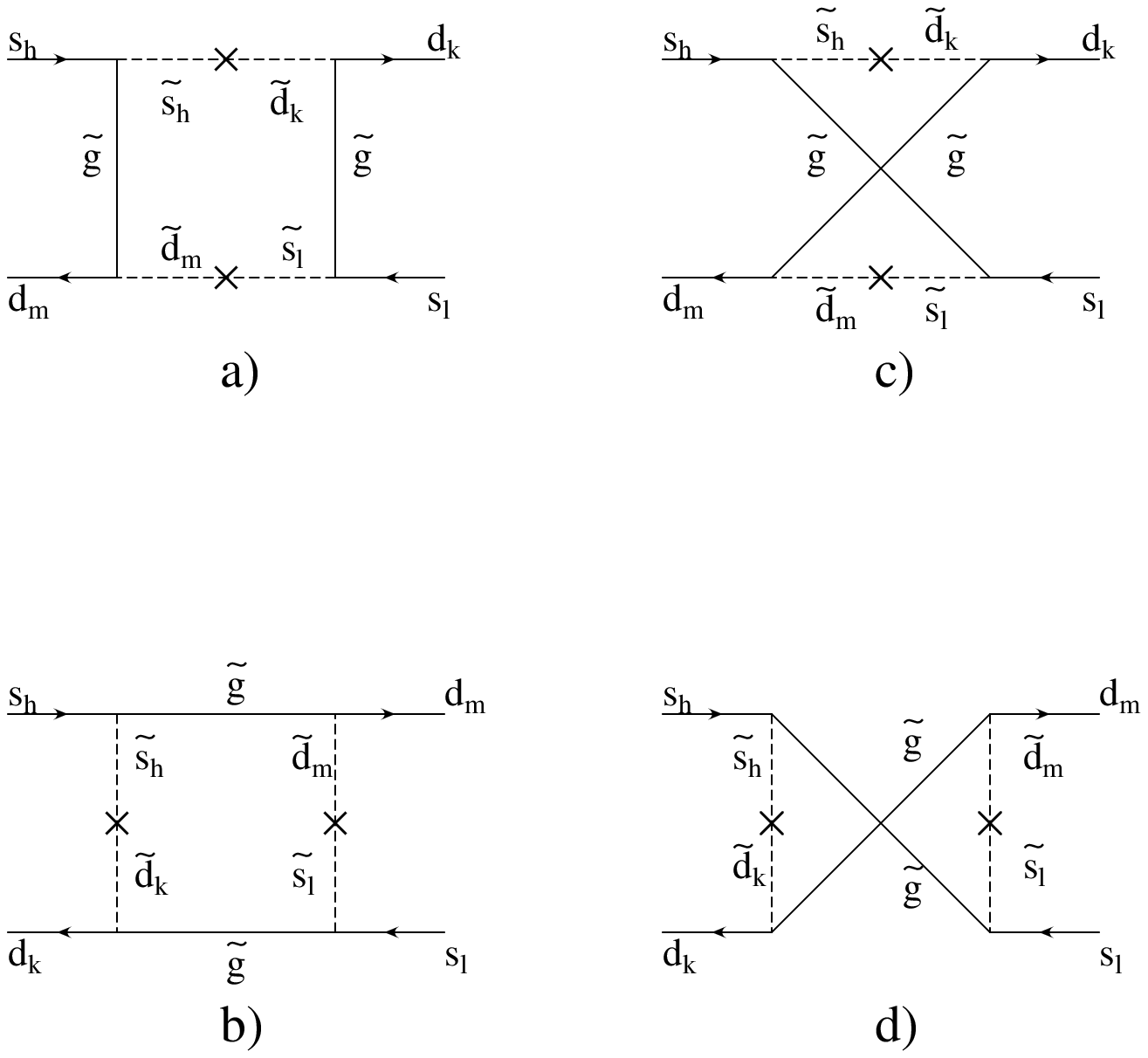} 
\caption{Feynman diagrams for $\Delta S=2$ transitions, with
  $h,k,l,m=\{L,R\}$.} 
\label{fig:ds2}
} 
Having calculated the amplitude from the diagrams in
fig.~\ref{fig:ds2}, one has to choose a basis of local operators and
perform the matching of the full theory to the one described by ${\cal
  H}_{\rm eff}^{\Delta S=2}$. We have adopted the form
\begin{equation}
  \label{eq:defheff}
  {\cal H}_{\rm eff}^{\Delta S=2}=\sum_{i=1}^{5} C_i\, Q_i +
  \sum_{i=1}^{3} \tilde{C}_i\, \tilde{Q}_i
\end{equation}
where
\begin{eqnarray}
        Q_1 & = & \db^{\alpha}_L \gd s^{\alpha}_L \db^{\beta}_{L} \gu
        s^{\beta}_L\; , 
        \nonumber \\
        Q_2 & = & \db^{\alpha}_R  s^{\alpha}_L \db^{\beta}_R s^{\beta}_L\; ,
        \nonumber \\
        Q_3 & = & \db^{\alpha}_R  s^{\beta}_L \db^{\beta}_R s^{\alpha}_L\; ,
        \nonumber \\
        Q_4 & = & \db^{\alpha}_R  s^{\alpha}_L \db^{\beta}_L s^{\beta}_R\; ,
        \nonumber \\
        Q_5 & = & \db^{\alpha}_R  s^{\beta}_L \db^{\beta}_L s^{\alpha}_R\; ,
        \label{Qi}
\end{eqnarray}
and the operators $\tilde{Q}_{1,2,3}$ are obtained from the
$Q_{1,2,3}$ by the exchange $ L \leftrightarrow R$.  Here
$q_{R,L}=P_{R,L}\,q$, with $P_{R,L}=(1 \pm \gamma_5)/2$, and $\alpha$
and $\beta$ are colour indices.

At the lowest order in QCD, we obtain 
the following results for the Wilson coefficients:
\begin{eqnarray}
        C_1&=&-\frac{\als^2}{216 m_{\tilde{q}}^2}  
        \left(  24\,x\,f_6(x) + 66\,\tilde{f}_6(x) \right)
        \left(\delta^d_{12}\right)^2_{LL} \,,
        \nonumber \\
        C_2&=&-\frac{\als^2}{216 m_{\tilde{q}}^2} 
         \, 204\,x\,f_6(x) \left(\delta^d_{12}\right)^2_{RL} \,,
        \nonumber \\
        C_3&=& \frac{\als^2}{216 m_{\tilde{q}}^2} 
        \,36\,x\,f_6(x) \left(\delta^d_{12}\right)^2_{RL} \,,
        \nonumber \\
        C_4&=&-\frac{\als^2}{216 m_{\tilde{q}}^2} 
        \biggl[  \left( 504\,x\,f_6(x) - 72\,\tilde{f}_6(x)\right)
        \left(\delta^d_{12}\right)_{LL}\left(\delta^d_{12}\right)_{RR}
        \nonumber \\
        &&- 132\,\tilde{f}_6(x)
        \left(\delta^d_{12}\right)_{LR}\left(\delta^d_{12}\right)_{RL}
        \biggr] \,,
        \nonumber \\
        C_5&=&-\frac{\als^2}{216 m_{\tilde{q}}^2} 
        \biggl[  \left( 24\,x\,f_6(x) + 120\,\tilde{f}_6(x)\right)
        \left(\delta^d_{12}\right)_{LL}\left(\delta^d_{12}\right)_{RR}
        \nonumber \\ 
        &&- 180\,\tilde{f}_6(x)
        \left(\delta^d_{12}\right)_{LR}\left(\delta^d_{12}\right)_{RL}
        \biggr] \,,
        \nonumber \\
        \tilde{C}_1&=& -\frac{\als^2}{216 m_{\tilde{q}}^2}  
        \left(  24\,x\,f_6(x) + 66\,\tilde{f}_6(x) \right)
        \left(\delta^d_{12}\right)^2_{RR} \,,
        \nonumber \\
        \tilde{C}_2&=&-\frac{\als^2}{216 m_{\tilde{q}}^2} 
         \, 204\,x\,f_6(x) \left(\delta^d_{12}\right)^2_{LR} \,,
        \nonumber \\
        \tilde{C}_3&=& \frac{\als^2}{216 m_{\tilde{q}}^2} 
        \,36\,x\,f_6(x) \left(\delta^d_{12}\right)^2_{LR} \,, 
        \label{inicoeff}
\end{eqnarray}
where $x=m^2_{\tilde{g}}/m_{\tilde{q}}^2$ and the functions $f_6(x)$ and 
$\tilde{f}_6(x)$ are given by:
\begin{eqnarray}
f_6(x)&=&\frac{6(1+3x)\ln x +x^3-9x^2-9x+17}{6(x-1)^5}\; , \nonumber  \\
\tilde{f}_6(x)&=&\frac{6x(1+x)\ln x -x^3-9x^2+9x+1}{3(x-1)^5}\; . 
\end{eqnarray}
In the absence of the $O(\alpha_s)$ corrections to the matching, we
interpret the $C_i$ given above as coefficients computed at the large
energy scale $M_S\sim m_{\tilde q} \sim m_{\tilde g}$, i.e. $C_i
\equiv C_i(M_S)$. 

The Next-to-Leading anomalous dimension matrix for the most general
${\cal H}_{\rm eff}^{\Delta F=2}$ has been recently computed
\cite{ciuchini}.  We use the Regularisation-Independent anomalous
dimension in the Landau gauge (LRI), since we will make use of matrix
elements computed in lattice QCD with the same choice of
renormalisation scheme (see ref.~\cite{ciuchini} for details on the
computation).

A full NLO computation would also require the $O(\alpha_s)$
corrections to the matching conditions in eq.~(\ref{inicoeff}).
Unfortunately, such corrections are not available yet. One might argue
that, being of order $\alpha_s(M_S)$, these contributions should be
small, as suggested by the cases of the SM and of the two Higgs
doublet model; however, this statement can only be confirmed by an
explicit computation.  Unfortunately, due to the absence of
$O(\alpha_s)$ corrections to the matching, our ${\cal H}_{\rm
  eff}^{\Delta F=2}$ will be affected by a residual scheme dependence,
which would be cancelled by the missing terms of order
$\alpha_s(M_S)$.

\TABLE{
\begin{tabular}{||c|c||}
        \hline \hline
        Constants & Values  \\
        \hline
        \hline
        $M_K$ & $497.67$ MeV  \\
        \hline
        $f_{K}$ & $159.8$ MeV  \\
        \hline
        $m_{d}(2 {\rm GeV})$ & $7$ MeV  \\
        \hline
        $m_{s}(2 {\rm GeV}) $ & $125$ MeV  \\
        \hline
        $m_{c}$ & $1.3$ GeV  \\
        \hline
        $m_{b}$ & $4.3$ GeV  \\
        \hline
        $m_t$ & $175$ GeV \\
        \hline
        $\als(M_{Z})$ & $0.119$ \\
        \hline \hline
\end{tabular}
\label{parameters}
\caption{Constants used in the phenomenological analysis.}
}

Eq.~(\ref{inicoeff}) is obtained by integrating out all
SUSY particles simultaneously. We then have to evolve the coefficients
down to the hadronic scale $\mu=2$ GeV, at which we have evaluated the
matrix elements.  The SM contribution can be computed
independently and evolved from $M_W$ to $\mu$ using the well-known NLO
QCD corrections \cite{buras}.

We give here an analytic formula for the expression of the Wilson
coefficients at the scale $\mu=2$ GeV as a function of the initial
conditions at the SUSY scale $C(M_S)$ and of $\alpha_s(M_S)$. This
formula has been obtained by using the values in
Table~\ref{parameters} for the SM parameters. 

For $M_S>m_t$ we obtain
\begin{equation}
\label{eq:magic1}
C_r(\mu)=\sum_i \sum_s  
             \left(b^{(r,s)}_i + \eta \,c^{(r,s)}_i\right) 
             \eta^{a_i} \,C_s(M_S),
\end{equation}
where, in the evolution of the coefficients from $M_S$, we have chosen
$M_S= (M_{\tilde g} + M_{\tilde q})/2$.
$\eta=\alpha_s(M_S)/\alpha_s(m_t)$, $\mu= 2$ GeV and the magic numbers
are given below:\footnote{In the previously published version of this paper, 
  \jhep{10}{1998}{008}, the magic numbers in eq.~(\ref{eq:magic2}) were 
  erroneously given in the basis in eq.~(13) of ref.~\cite{ciuchini} instead
  of the basis (\ref{Qi}) used in this work (we thank P.~Slavich and 
  F.~Zwirner for pointing this out to us). However, all the
  numerical results in Sect.~\ref{sec:numerics} and \ref{sec:R} of the 
  previous version were correct, since they were obtained using the full 
  evolution matrix instead of eq.~(\ref{eq:magic1}).}
\begin{equation}
\label{eq:magic2}
\begin{array}{l l}
a_i=(0.29,-0.69,0.79,-1.1,0.14)& \\ 
& \\ 
b^{(11)}_i=(0.82,0,0,0,0),& 
c^{(11)}_i=(-0.016,0,0,0,0),\\ 
b^{(22)}_i=(0,2.4,0.011,0,0),& 
c^{(22)}_i=(0,-0.23,-0.002,0,0),\\ 
b^{(23)}_i=(0,-0.63,0.17,0,0),& 
c^{(23)}_i=(0,-0.018,0.0049,0,0),\\ 
b^{(32)}_i=(0,-0.019,0.028,0,0),& 
c^{(32)}_i=(0,0.0028,-0.0093,0,0),\\ 
b^{(33)}_i=(0,0.0049,0.43,0,0),& 
c^{(33)}_i=(0,0.00021,0.023,0,0),\\ 
b^{(44)}_i=(0,0,0,4.4,0),& 
c^{(44)}_i=(0,0,0,-0.68,0.0055),\\ 
b^{(45)}_i=(0,0,0,1.5,-0.17),& 
c^{(45)}_i=(0,0,0,-0.35,-0.0062),\\ 
b^{(54)}_i=(0,0,0,0.18,0),& 
c^{(54)}_i=(0,0,0,-0.026,-0.016),\\ 
b^{(55)}_i=(0,0,0,0.061,0.82),& 
c^{(55)}_i=(0,0,0,-0.013,0.018),\\ 
\end{array}
\end{equation}
and we have only written the non-vanishing entries. The magic numbers
for the evolution of $\tilde C_{1-3}$ are the same as the ones for the
evolution of $C_{1-3}$.  Formulae~(\ref{eq:magic1}) and
(\ref{eq:magic2}) can be used in connection with the $B$-parameters
evaluated at $\mu=2$ GeV, given in eq.~(\ref{eq:fres}), to compute the
contribution to $\Delta M_K$ and $\varepsilon_K$ at the NLO in QCD for
any model of new physics in which the new contributions with respect
to the SM originate from extra heavy particles. One just has to plug
in the expression for the $C_i$ evaluated at the large energy scale
$M_S$ in his favourite model. When the $O(\alpha_s)$ corrections to
the $C_i(M_S)$ are available, one can obtain a full NLO,
regularisation-independent result; in the cases where this corrections
have not been computed yet, the results contain a residual systematic
uncertainty of order $\alpha_s(M_S)$. We note that, due to the
presence of large entries in the NLO anomalous dimension matrix, a
systematic uncertainty of a few percents is present in the QCD
evolution from the SUSY scale to the hadronic one.  For this reason we
only present the significant figures for the magic numbers.

\section{Hadronic Matrix Elements}
\label{sec:ME}

The matrix elements of the operators $Q_i$ between $K$ mesons in the
VIA are given by:
\begin{eqnarray}
        \langle K^{0} \vert Q_{1} 
        \vert \bar{K}^{0} \rangle_{\rm VIA} & = & 
        \frac{1}{3}M_Kf_{K}^{2}\; ,
        \nonumber \\
        \langle K^{0} \vert Q_{2} 
        \vert \bar{K}^{0} \rangle_{\rm VIA}  & = & -\frac{5}{24}  
        \left(\frac{M_K}{m_{s}+m_{d}}\right)^{2}M_Kf_{K}^{2}\; ,
        \nonumber \\
        \langle K^{0} \vert Q_{3} 
        \vert \bar{K}^{0} \rangle_{\rm VIA}  & = & \frac{1}{24} 
        \left(\frac{M_K}{m_{s}+m_{d}}\right)^{2}M_Kf_{K}^{2}\; ,
        \nonumber \\
        \langle K^{0} \vert Q_{4} 
        \vert \bar{K}^{0} \rangle_{\rm VIA} & = & \left[\frac{1}{24} +
        \frac{1}{4} 
        \left(\frac{M_K}{m_{s}+m_{d}}\right)^{2}\right]M_Kf_{K}^{2}\; ,
        \nonumber \\
        \langle K^{0} \vert Q_{5}
        \vert \bar{K}^{0} \rangle_{\rm VIA} & = &\left[\frac{1}{8} +
        \frac{1}{12} 
        \left(\frac{M_K}{m_{s}+m_{d}}\right)^{2}\right] M_Kf_{K}^{2}\; ,
        \label{me}
\end{eqnarray}
where $M_K$ is the mass of the $K^0$ meson and $m_{s}$ and $m_{d}$ are
the masses of $s$ and $d$ quarks respectively. Here and in the
following, the same expressions of the $B$-parameters of the operators
$Q_{1-3}$ are valid for the operators $\tilde Q_{1-3}$, since strong
interactions preserve parity. 

In the case of the renormalised operators, we define the
$B$-parameters as follows:
\bea 
\langle \bar K^{0} \vert \hat Q_{1} (\mu) 
\vert K^{0} \rangle & = & 
\frac{1}{3}M_Kf_{K}^{2} B_1(\mu)\; ,
\nonumber \\
\langle
\bar K^{0} \vert \hat Q_{2} (\mu) \vert K^{0} \rangle &=& -\frac{5}{24}
\left( \frac{ M_K }{ m_{s}(\mu) + m_d(\mu) }\right)^{2}
M_K f_{K}^{2} B_{2}(\mu) \nonumber  \\
\langle \bar K^{0} \vert \hat Q_{3} (\mu) \vert K^{0} \rangle &=&
\frac{1}{24} \left( \frac{ M_K }{ m_{s}(\mu) + m_d(\mu) }\right)^{2}
M_K f_{K}^{2} B_{3}(\mu) \nn \\
\langle \bar K^{0} \vert \hat Q_{4} (\mu) \vert K^{0} \rangle &=& \frac{1}{4}
\left( \frac{ M_K }{ m_{s}(\mu) + m_d(\mu) }\right)^{2}
M_K f_{K}^{2} B_{4}(\mu) \nn \\
\langle \bar K^{0} \vert \hat Q_{5} (\mu) \vert K^{0} \rangle &=&
\frac{1}{12} \left( \frac{ M_K }{ m_{s}(\mu) + m_d(\mu) }\right)^{2}
M_K f_{K}^{2} B_{5}(\mu) \label{eq:bpars} \ , 
\eea 
where the notation $\hat Q_{i}(\mu)$ (or simply $\hat Q_{i}$) denotes
the operators renormalised at the scale $\mu$.  

A few words of explanation are necessary at this point. The
$B$-parameter of the matrix element $\langle \bar K^0 \vert Q_1 \vert
K^0 \rangle$, commonly known as $B_K$, has been extensively studied on
the lattice due to its phenomenological relevance~\cite{sharpe:lat96},
and used in many phenomenological studies~\cite{parodi}. For the other
operators, instead, all the phenomenological analyses beyond the SM
have used $B$-parameters equal to one, which in some cases, as will be
shown below, is a very crude approximation.

In eq.~(\ref{eq:bpars}) the operators and the quark masses are
renormalised in the same scheme (RI, $\overline{MS}$, etc.)  at the
scale $\mu$ and the numerical results for the $B$-parameters,
$B_{i}(\mu)$, presented below refer to the Landau RI scheme.
Moreover, without loss of generality, we have omitted terms which are
of higher order in the chiral expansion, see eqs.~(\ref{me}), and
which are usually included in the definition of the $B$-parameters
(see ref.~\cite{roma} for a thorough discussion of the advantages of
our definition of the $B$-parameters).

In our numerical study, we have used, for $\mu=2$ GeV:
\bea
B_{1}(\mu) &=& 0.60(6)  \label{eq:fres} \\ 
B_{2}(\mu) &=& 0.66(4)  \nn \\
B_{3}(\mu) &=& 1.05(12) \nn \\
B_{4}(\mu) &=& 1.03(6)  \nn \\
B_{5}(\mu) &=& 0.73(10) \nn 
\eea

The central value we use for $B_1=B_K$ corresponds to $B_K^{\overline{
  MS}}(2\, {\rm GeV})=0.61$, in agreement with the recent estimates of
refs.~\cite{sharpe:lat96,gupta}. $B_{2-5}$ have been taken from
ref.~\cite{roma}, where all details of the computation can be found
(for another determination of these $B$-parameters, calculated with
perturbative renormalization, see ref.~\cite{otherb}).

\section{Numerical Analysis of $\Delta S =2$ Processes}
\label{sec:numerics}

We now present the results of a model-independent analysis of 
$K^{0} $--$\bar{K}^{0}$ mixing, for the case $m_{\tilde{q}} \sim
m_{\tilde{g}}$.
The $K_{L}$--$K_{S}$ mass difference $\Delta M_K$ and the CP-violating
parameter $\varepsilon_K$ are given by:
\begin{eqnarray}
        &&\Delta M_K = 2 \Re \langle K^{0} \vert {\cal
        H}_{\rm eff}^{\Delta S=2} 
        \vert \bar{K}^{0} \rangle \;,
        \label{defdmk}\\
        &&\varepsilon = \frac{1}{\sqrt{2} \Delta M_K}
                          \Im  \langle K^{0} \vert {\cal H}_{\rm
                          eff}^{\Delta S=2}   
        \vert \bar{K}^{0} \rangle \;.
        \label{defepsi}
\end{eqnarray}

The SUSY (gluino-mediated) contribution to the low-energy ${\cal
  H}_{\rm eff}^{\Delta S=2}$ contains two real and four complex
unknown parameters: $m_{\tilde q}$, $m_{\tilde g}$,
$\left(\delta^d_{12}\right)_{LL}$, $\left(\delta^d_{12}\right)_{LR}$,
$\left(\delta^d_{12}\right)_{RL}$ and
$\left(\delta^d_{12}\right)_{RR}$.  These parameters can be determined
once a specific SUSY model is chosen, and the NLO contribution to
$\Delta M_K$ and $\varepsilon_K$ can be directly computed by using the
expression for the low-energy ${\cal H}_{\rm eff}^{\Delta S=2}$ and the
$B$-parameters given in eqs.~(\ref{inicoeff}), (\ref{eq:magic1}) and
(\ref{eq:fres}).

However, it is also useful to provide a set of model-independent
constraints on the individual $\delta$-parameters, obtained by
neglecting the interference between the different SUSY contributions.
This is justified (a posteriori) by noting that the constraints on
different $\delta$-parameters in the Kaon case exhibit a hierarchical
structure, and therefore interference effects between different
contributions would require a large amount of fine tuning.

The model-independent constraints are obtained by imposing that the
sum of the SUSY contributions proportional to a single
$\delta$-parameter and of the SM contributions to $\Delta M_K$ and
$\varepsilon_K$ does not exceed the experimental value for these
quantities.

A comment on the SM contributions is necessary at this point.  Let us
first consider $\Delta M_K$. The short-distance SM contribution to
this process is dominated by charm-quark exchange. Hence its value can
be computed once the CKM elements $V_{cd}$ and $V_{cs}$ are
determined. Making the very reasonable hypothesis that SUSY
contributions to tree-level weak decays are negligible, the
determination of the CKM elements $V_{cd}$ and $V_{cs}$ is unaffected
by SUSY, and therefore the SM contribution to $\Delta M_K$ can be
computed even in the presence of large SUSY contributions to loop
processes. We are not considering here long-distance contributions to
$\Delta M_K$. The constraints we obtain from $\Delta M_K$ are
therefore more conservative if the long-distance SM contributions add
up to the short-distance ones.

On the other hand, the SM contribution to $\varepsilon_K$ depends on
the phase in the CKM matrix. This phase is usually extracted, in the
context of the SM, from the analysis of $\varepsilon_K$ and $B^0$--$
\bar B^0$ mixing. However, in the presence of large (unknown) SUSY
contributions to $K^0 $--$\bar K^0$ and $B^0 $--$\bar B^0$ mixing, the
extraction of the CKM phase is not possible. We therefore treat it as
a free parameter. Since the SM contribution to $\varepsilon_K$ is
always positive, and vanishes for vanishing CKM phase, in order to
obtain a conservative limit on SUSY contributions we set the CKM
phase to zero, and allow the SUSY contribution to saturate the
experimental value of $\varepsilon_K$.

We make use of the $B$-parameters given in eq.~(\ref{eq:fres}) and
subtract one standard deviation to their central values, in order to
extract a more conservative bound on SUSY parameters.

Barring accidental cancellations, following the procedure explained
above, we obtain the limits on the $\delta$-parameters reported in
Tables~\ref{tab:reds2_200}--\ref{tab:imds2_1000}.  We give results for
different values of the average squark mass, since the naive scaling
of the constraints like $1/\tilde{m}$ is modified by the perturbative
QCD corrections.

For Left-Right mass insertions, we consider two possible (extreme)
cases: $|(\delta^{d}_{12})_{LR}|\gg |(\delta^{d}_{12})_{RL}|$ and
$(\delta^{d}_{12})_{LR} = (\delta^{d}_{12})_{RL}$. In the second case,
we combine the contributions proportional to
$(\delta^{d}_{12})^2_{LR}$, $(\delta^{d}_{12})^2_{RL}$ and
$(\delta^{d}_{12})_{LR} (\delta^{d}_{12})_{RL}$. This approach
improves the one of refs.~\cite{hagelin} and \cite{bagger}, where
contributions proportional to $(\delta^{d}_{12})_{LR}
(\delta^{d}_{12})_{RL}$ where considered independently from the ones
proportional to $(\delta^{d}_{12})^2_{LR}$ and
$(\delta^{d}_{12})^2_{RL}$. The interference between the various
Left-Right terms, in the case $(\delta^{d}_{12})_{LR} =
(\delta^{d}_{12})_{RL}$, produces a cancellation effect which is
particularly sizeable at the NLO for $m_{\tilde{q}}=500$ GeV, around
$x=m^2_{\tilde{g}}/m^2_{\tilde{q}}=1$, as one can see from
Tables~\ref{tab:reds2_500} and \ref{tab:imds2_500}. This means that
our results are not reliable for this particular case, since
interference effects can change dramatically the final results for
small variations of the input parameters.

On the other hand, the $(\delta^{d}_{12})_{LL} (\delta^{d}_{12})_{RR}$
contribution can be treated independently from the
$(\delta^{d}_{12})^2_{LL}$ and $(\delta^{d}_{12})^2_{RR}$ ones, since
the latter ones do not generate Left-Right operators and are therefore
suppressed with respect to the first contribution.

Let us compare our results with the previous analyses of
refs.~\cite{GGMS} and \cite{bagger}. In both papers, the matrix
elements were computed in the VIA, whereas we are using the
$B$-parameters computed in lattice QCD. This allows us a consistent
matching of the renormalisation-scheme and $\mu$ dependence of the
QCD-corrected ${\cal H}_{\rm eff}^{\Delta S=2}$. Moreover we keep
into account important non-perturbative effects present in the
hadronic matrix elements.

If we use the same input parameters as in ref.~\cite{GGMS}, and
neglect the SM contribution, we reproduce the results of
ref.~\cite{GGMS} for the QCD-uncorrected case. The values quoted in
Tables \ref{tab:reds2_500}-\ref{tab:imds2_500} differ from the ones of
ref.~\cite{GGMS} due to the inclusion of the SM contributions in the
present analysis and to the different choice of input parameters.  As
one can read from Table \ref{tab:reds2_500}, our QCD-uncorrected and
LO results are also different from those of ref.~\cite{bagger}. This
is due to the different choice of the input parameters, to the neglect
of the SM contribution in ref.~\cite{bagger} and to the choice of the
scale $\mu$. In particular, the results in Table \ref{tab:reds2_500}
correspond to $\mu=2$ GeV, while the authors of ref.~\cite{bagger}
chose to evolve the ${\cal H}_{\rm eff}^{\Delta S=2}$ down to the
scale $\mu^\prime$ defined by $\alpha_s(\mu^\prime)=1$. This choice
may be questionable, since perturbation theory is expected to break
down at such low scales.

\section{Constraints on R-parity violating couplings}
\label{sec:R}

In this section we update the constraints that can be derived from
$\Delta S=2$ processes on some R-parity violating couplings.

If R-parity invariance is not imposed, 
the superpotential may contain the following additional terms:
\be
\lambda^{\prime \prime}_{ijk}u^c_{i}d^c_{j}d_{k}^c +
\lambda^{\prime}_{ijk}L_{i}Q_{j}d_{k}^c + \lambda_{ijk}L_{i}L_{j}e_{k}^c \, ,
\label{superp}
\ee where $i$, $j$ and $k$ are generation indices, and we omitted
possible bilinear R-parity violating terms \cite{rpar}.  The
$\lambda_{ijk}$ and $\lambda^{\prime}_{ijk}$ are lepton number
violating Yukawa couplings, while the $\lambda^{\prime \prime}_{ijk}$
are baryon number violating ones. The $\lambda_{ijk}$ are
antisymmetric under the exchange of the first two generation indices,
while the $\lambda^{\prime \prime}_{ijk}$ are antisymmetric under the
interchange of the last two flavour indices.  We consider the
R-violating couplings in the super-CKM basis.

Stringent constraints can be imposed on R-violating couplings by
considering a variety of processes (see ref.~\cite{rrew} for a recent
review on this subject). We concentrate on the limits that can be
obtained from $\Delta S=2$ processes, and in particular on the ones
arising from Left-Right effective operators.

The most stringent constraints arise from the tree-level sneutrino-mediated
contributions, that have the form
\begin{equation}
  \label{eq:tree}
  C_4(m_{\tilde \nu})=\sum_i \frac{\lambda^{\prime}_{i21}
  \lambda^{\prime^*}_{i12} }{m_{\tilde \nu_i}^2}\, .
\end{equation}

Additional constraints can be obtained by considering box diagrams
with four R-violating couplings. As was pointed out in
ref.~\cite{pisa}, however, stronger constraints on the product of two
$\lambda^\prime$ couplings can be obtained by considering box diagrams
with the exchange of a slepton, together with a $W$ or a $H^\pm$. We
consider only contributions with top-quark exchange, since when light
quarks are present in the loop the OPE is modified and the QCD
evolution is much more involved. This computation is not available
yet. In eq.~(3) of ref.~\cite{pisa}, the coefficient of the effective
Hamiltonian at the electroweak scale was computed for
$m_{H^\pm}=m_{\tilde l}$, by integrating out simultaneously all heavy
particles. In order to avoid the appearance of large logs of the form
$\ln m_t^2/m_{\tilde l}^2$ in the matching conditions, we consider
this contribution for $m_{H^\pm}=m_{\tilde l}=200$ GeV, and derive
constraints on the product of the R-violating couplings
$\lambda^{\prime}_{i31} \lambda^{\prime^*}_{i32}$.

Using the same procedure as that followed for R-conserving SUSY
contributions, we obtain the constraints given in Table
\ref{tab:rpar}. As in the previous Section, we put the CKM phase to
zero in order to obtain more conservative limits.
 
\section{Conclusions}
\label{sec:concl}

In our work we have provided an improved computation of the
gluino-mediated SUSY contributions to $K^0 $--$\bar K^0$ mixing in
the framework of the mass insertion method. The improvement consists
in introducing the NLO QCD corrections to the ${\cal H}_{\rm
  eff}^{\Delta S=2}$ \cite{ciuchini} and in replacing the VIA
$B$-parameters with their recent lattice computation \cite{roma}. As a
glimpse at Tables \ref{tab:reds2_200}--\ref{tab:imds2_1000} readily
reveals, these improvements affect previous results in a different
way, according to the operators of ${\cal H}_{\rm eff}^{\Delta S=2}$
one considers. The effect is particularly large for Left-Right
operators.

We have also improved in the same way the constraints on some products
of R-violating $\lambda^\prime$-type couplings that can be derived
from $K^0 $--$\bar K^0$ mixing. Here the effects of QCD corrections, that
were never considered before, are also very large, due to the
Left-Right structure of the operator involved.

We have provided an analytic formula for the most general low-energy
${\cal H}_{\rm eff}^{\Delta S=2}$ at the NLO, in terms of the Wilson
coefficients at the high energy scale. This formula can be readily
used to compute $\Delta M_K$ and $\varepsilon_K$ in any extension of
the SM with new heavy particles.

The above results suggest several improvements for theoretical
analyses. As for $K$-mixing, a computation of the Wilson
coefficients at $M_S$ is needed to have the complete NLO QCD-corrected
and scheme-independent result. Another issue that needs to be
clarified is the role played by the other classes of FCNC
contributions which are present in SUSY in addition to the gluino
exchange. In particular the chargino-squark contributions should be
considered, although, as we said, for generic squark mass matrices, we
expect the gluino exchange to set the correct size of the whole FCNC
contributions.

Even more interesting is to extend this analysis to $\Delta B=2$
processes and to combine the limits which can be derived by
considering FCNC processes involving $B$ and $K$ mesons
simultaneously. The lattice derivation of the $B_B$ parameters is in
progress and so we are confident to be able to perform a similar
analysis in the $B$ case.

FCNC and CP violating phenomena (in particular in $B$ physics) are
promising candidates for some indirect SUSY signal before LHC, and are
in many ways complementary to direct SUSY searches. From this point of
view the theoretical effort to improve as much as possible our
precision on FCNC computations in a SUSY model-independent framework
is certainly worth and, hopefully, rewarding.

\section*{Acknowledgements}
We thank Guido Michelon for discussions and collaboration at the early
stages of this work. M.C. thanks the CERN TH Division for the
hospitality during the completion of this work. L.S. acknowledges the
support of the German Bundesministerium f{\"u}r Bildung und Forschung
under contract 06 TM 874 and DFG Project Li 519/2-2. The work of A.M.
was partly supported by the TMR project ``Beyond the Standard Model''
under contract number ERBFMRX CT96 0090. L.C., V.L., G.M., I.S. and
A.V. acknowledge partial support by the MURST. M.T. acknowledges the
support of PPARC through grant GR/L22744.


 \TABLE{ 
 \begin{tabular}{||c|c|c|c|c||}  \hline \hline 
 & NO QCD, VIA & LO, VIA & LO, Lattice $B_i$ & NLO, Lattice $B_i$  \\ 
 \hline 
 $x$ & \multicolumn{4}{c||}{$\sqrt{|\Re  (\delta^{d}_{12})_{LL}^{2}|} $} \\ 
\hline 
 0.3& $5.0\times 10^{-3}$ 

& $5.7\times 10^{-3}$ 

& $7.7\times 10^{-3}$ 

& $7.7\times 10^{-3}$ 
\\ 
1.0& $1.1\times 10^{-2}$ 

& $1.2\times 10^{-2}$ 

& $1.6\times 10^{-2}$ 

& $1.6\times 10^{-2}$ 
\\ 
4.0& $2.5\times 10^{-2}$ 

& $2.9\times 10^{-2}$ 

& $3.9\times 10^{-2}$ 

& $3.9\times 10^{-2}$ 
\\ 
\hline  
 $x$ & \multicolumn{4}{c||}{$\sqrt{|\Re  (\delta^{d}_{12} )_{LR}^{2}|} 
 \qquad (|(\delta^{d}_{12})_{LR}|\gg 
|(\delta^{d}_{12})_{RL}|)$} \\ 
\hline 
 0.3& $1.1\times 10^{-3}$ 
 
& $8.4\times 10^{-4}$ 
 
& $1.1\times 10^{-3}$ 
 
& $9.6\times 10^{-4}$ 
\\ 
1.0& $1.2\times 10^{-3}$ 
 
& $9.3\times 10^{-4}$ 
 
& $1.2\times 10^{-3}$ 
 
& $1.1\times 10^{-3}$ 
\\ 
4.0& $1.8\times 10^{-3}$ 
 
& $1.3\times 10^{-3}$ 
 
& $1.6\times 10^{-3}$ 
 
& $1.5\times 10^{-3}$ 
\\ 
\hline  
 $x$ & \multicolumn{4}{c||}{$\sqrt{|\Re  (\delta^{d}_{12} )_{LR}^{2}|} 
\qquad ((\delta^{d}_{12})_{LR} = 
 (\delta^{d}_{12} )_{RL})$} \\ 
 \hline 
 0.3& $2.0\times 10^{-3}$ 
 
& $1.4\times 10^{-3}$ 
 
& $8.9\times 10^{-4}$ 
 
& $6.7\times 10^{-4}$ 
\\ 
1.0& $1.1\times 10^{-3}$ 
 
& $9.7\times 10^{-4}$ 
 
& $1.8\times 10^{-3}$ 
 
& $3.0\times 10^{-3}$ 
\\ 
4.0& $1.3\times 10^{-3}$ 
 
& $1.0\times 10^{-3}$ 
 
& $1.4\times 10^{-3}$ 
 
& $1.3\times 10^{-3}$ 
\\ 
\hline  
 $x$ & \multicolumn{4}{c||}{$\sqrt{|\Re  (\delta^{d}_{12} )_{LL}
(\delta^{d}_{12})_{RR}|} $} \\ 
 \hline 
0.3& $6.4\times 10^{-4}$ 
 
& $3.9\times 10^{-4}$ 
 
& $4.0\times 10^{-4}$ 
 
& $3.3\times 10^{-4}$ 
\\ 
1.0& $7.1\times 10^{-4}$ 
 
& $4.4\times 10^{-4}$ 
 
& $4.5\times 10^{-4}$ 
 
& $3.7\times 10^{-4}$ 
\\ 
4.0& $1.0\times 10^{-3}$ 
 
& $6.1\times 10^{-4}$ 
 
& $6.2\times 10^{-4}$ 
 
& $5.2\times 10^{-4}$ 
\\ 
\hline  \hline 
 \end{tabular} 
 \caption{Limits on $\mbox{Re}\left(\delta_{ij} 
 \right)_{AB}\left(\delta_{ij}\right)_{CD}$, with $A,B,C,D=(L,R)$, for an
  average squark mass $m_{\tilde{q}}=200$ GeV and for different values
 of $x=m_{\tilde{g}}^2/m_{\tilde{q}}^2$.} 
 \label{tab:reds2_200} 
} 
 \TABLE{ 
 \begin{tabular}{||c|c|c|c|c||}  \hline \hline 
 & NO QCD, VIA & LO, VIA & LO, Lattice $B_i$ & NLO, Lattice $B_i$  \\ 
 \hline 
 $x$ & \multicolumn{4}{c||}{$\sqrt{|\Im  (\delta^{d}_{12} )_{LL}^{2}|} $} \\ 
\hline 
0.3& $6.7\times 10^{-4}$ 
 
& $7.5\times 10^{-4}$ 
 
& $1.0\times 10^{-3}$ 
 
& $1.0\times 10^{-3}$ 
\\ 
1.0& $1.4\times 10^{-3}$ 
 
& $1.6\times 10^{-3}$ 
 
& $2.2\times 10^{-3}$ 
 
& $2.2\times 10^{-3}$ 
\\ 
4.0& $3.3\times 10^{-3}$ 
 
& $3.8\times 10^{-3}$ 
 
& $5.1\times 10^{-3}$ 
 
& $5.1\times 10^{-3}$ 
\\ 
\hline  
 $x$ & \multicolumn{4}{c||}{$\sqrt{|\Im  (\delta^{d}_{12} )_{LR}^{2}|} 
\qquad (|(\delta^{d}_{12})_{LR}|\gg
 |(\delta^{d}_{12})_{RL}|)$} \\ 
 \hline 
0.3& $1.5\times 10^{-4}$ 
 
& $1.1\times 10^{-4}$ 
 
& $1.4\times 10^{-4}$ 
 
& $1.3\times 10^{-4}$ 
\\ 
1.0& $1.6\times 10^{-4}$ 
 
& $1.2\times 10^{-4}$ 
 
& $1.6\times 10^{-4}$ 
 
& $1.4\times 10^{-4}$ 
\\ 
4.0& $2.3\times 10^{-4}$ 
 
& $1.7\times 10^{-4}$ 
 
& $2.2\times 10^{-4}$ 
 
& $1.9\times 10^{-4}$ 
\\ 
\hline  
 $x$ & \multicolumn{4}{c||}{$\sqrt{|\Im  (\delta^{d}_{12} )_{LR}^{2}|} 
\qquad ((\delta^{d}_{12})_{LR} = 
 (\delta^{d}_{12} )_{RL})$} \\ 
 \hline 
0.3& $2.6\times 10^{-4}$ 
 
& $1.9\times 10^{-4}$ 
 
& $1.2\times 10^{-4}$ 
 
& $8.9\times 10^{-5}$ 
\\ 
1.0& $1.5\times 10^{-4}$ 
 
& $1.3\times 10^{-4}$ 
 
& $2.4\times 10^{-4}$ 
 
& $3.9\times 10^{-4}$ 
\\ 
4.0& $1.8\times 10^{-4}$ 
 
& $1.4\times 10^{-4}$ 
 
& $1.8\times 10^{-4}$ 
 
& $1.7\times 10^{-4}$ 
\\ 
\hline  
 $x$ & \multicolumn{4}{c||}{$\sqrt{|\Im  (\delta^{d}_{12} )_{LL}
(\delta^{d}_{12})_{RR}|}$}  \\ 
 \hline 
0.3& $8.4\times 10^{-5}$ 
 
& $5.2\times 10^{-5}$ 
 
& $5.3\times 10^{-5}$ 
 
& $4.4\times 10^{-5}$ 
\\ 
1.0& $9.4\times 10^{-5}$ 
 
& $5.8\times 10^{-5}$ 
 
& $5.9\times 10^{-5}$ 
 
& $4.9\times 10^{-5}$ 
\\ 
4.0& $1.3\times 10^{-4}$ 
 
& $8.1\times 10^{-5}$ 
 
& $8.2\times 10^{-5}$ 
 
& $6.8\times 10^{-5}$ 
\\ 
\hline  \hline 
 \end{tabular} 
 \caption{Limits on $\mbox{Im}\left(\delta_{ij}
 \right)_{AB}\left(\delta_{ij}\right)_{CD}$, with $A,B,C,D=(L,R)$, 
 for an average squark mass $m_{\tilde{q}}=200$ GeV and for different values 
 of $x=m_{\tilde{g}}^2/m_{\tilde{q}}^2$.} 
 \label{tab:imds2_200} 
} 

 \TABLE{ 
 \begin{tabular}{||c|c|c|c|c||}  \hline \hline 
 & NO QCD, VIA & LO, VIA & LO, Lattice $B_i$ & NLO, Lattice $B_i$  \\ 
 \hline 
 $x$ & \multicolumn{4}{c||}{$\sqrt{|\Re  (\delta^{d}_{12})_{LL}^{2}|} $} \\ 
\hline 
 0.3& $1.4\times 10^{-2}$ 

& $1.6\times 10^{-2}$ 

& $2.2\times 10^{-2}$ 

& $2.2\times 10^{-2}$ 
\\ 
1.0& $3.0\times 10^{-2}$ 

& $3.4\times 10^{-2}$ 

& $4.6\times 10^{-2}$ 

& $4.6\times 10^{-2}$ 
\\ 
4.0& $7.0\times 10^{-2}$ 

& $8.0\times 10^{-2}$ 

& $1.1\times 10^{-1}$ 

& $1.1\times 10^{-1}$ 
\\ 
\hline  
 $x$ & \multicolumn{4}{c||}{$\sqrt{|\Re  (\delta^{d}_{12} )_{LR}^{2}|} 
 \qquad (|(\delta^{d}_{12})_{LR}|\gg 
|(\delta^{d}_{12})_{RL}|)$} \\ 
\hline 
 0.3& $3.1\times 10^{-3}$ 
 
& $2.3\times 10^{-3}$ 
 
& $2.8\times 10^{-3}$ 
 
& $2.6\times 10^{-3}$ 
\\ 
1.0& $3.4\times 10^{-3}$ 
 
& $2.5\times 10^{-3}$ 
 
& $3.1\times 10^{-3}$ 
 
& $2.8\times 10^{-3}$ 
\\ 
4.0& $4.9\times 10^{-3}$ 
 
& $3.5\times 10^{-3}$ 
 
& $4.4\times 10^{-3}$ 
 
& $3.9\times 10^{-3}$ 
\\ 
\hline  
 $x$ & \multicolumn{4}{c||}{$\sqrt{|\Re  (\delta^{d}_{12} )_{LR}^{2}|} 
\qquad ((\delta^{d}_{12})_{LR} = 
 (\delta^{d}_{12} )_{RL})$} \\ 
 \hline 
 0.3& $5.5\times 10^{-3}$ 
 
& $3.3\times 10^{-3}$ 
 
& $2.2\times 10^{-3}$ 
 
& $1.7\times 10^{-3}$ 
\\ 
1.0& $3.1\times 10^{-3}$ 
 
& $2.7\times 10^{-3}$ 
 
& $5.5\times 10^{-3}$ 
 
& $2.8\times 10^{-2}$ 
\\ 
4.0& $3.7\times 10^{-3}$ 
 
& $2.8\times 10^{-3}$ 
 
& $3.8\times 10^{-3}$ 
 
& $3.5\times 10^{-3}$ 
\\ 
\hline  
 $x$ & \multicolumn{4}{c||}{$\sqrt{|\Re  (\delta^{d}_{12} )_{LL}
(\delta^{d}_{12})_{RR}|} $} \\ 
 \hline 
0.3& $1.8\times 10^{-3}$ 
 
& $1.0\times 10^{-3}$ 
 
& $1.0\times 10^{-3}$ 
 
& $8.6\times 10^{-4}$ 
\\ 
1.0& $2.0\times 10^{-3}$ 
 
& $1.1\times 10^{-3}$ 
 
& $1.2\times 10^{-3}$ 
 
& $9.6\times 10^{-4}$ 
\\ 
4.0& $2.8\times 10^{-3}$ 
 
& $1.6\times 10^{-3}$ 
 
& $1.6\times 10^{-3}$ 
 
& $1.3\times 10^{-3}$ 
\\ 
\hline  \hline 
 \end{tabular} 
 \caption{Limits on $\mbox{Re}\left(\delta_{ij} 
 \right)_{AB}\left(\delta_{ij}\right)_{CD}$, with $A,B,C,D=(L,R)$, for an
  average squark mass $m_{\tilde{q}}=500$ GeV and for different values
 of $x=m_{\tilde{g}}^2/m_{\tilde{q}}^2$.} 
 \label{tab:reds2_500} 
} 
 \TABLE{ 
 \begin{tabular}{||c|c|c|c|c||}  \hline \hline 
 & NO QCD, VIA & LO, VIA & LO, Lattice $B_i$ & NLO, Lattice $B_i$  \\ 
 \hline 
 $x$ & \multicolumn{4}{c||}{$\sqrt{|\Im  (\delta^{d}_{12} )_{LL}^{2}|} $} \\ 
\hline 
0.3& $1.8\times 10^{-3}$ 
 
& $2.1\times 10^{-3}$ 
 
& $2.9\times 10^{-3}$ 
 
& $2.9\times 10^{-3}$ 
\\ 
1.0& $3.9\times 10^{-3}$ 
 
& $4.5\times 10^{-3}$ 
 
& $6.1\times 10^{-3}$ 
 
& $6.1\times 10^{-3}$ 
\\ 
4.0& $9.2\times 10^{-3}$ 
 
& $1.1\times 10^{-2}$ 
 
& $1.4\times 10^{-2}$ 
 
& $1.4\times 10^{-2}$ 
\\ 
\hline  
 $x$ & \multicolumn{4}{c||}{$\sqrt{|\Im  (\delta^{d}_{12} )_{LR}^{2}|} 
\qquad (| (\delta^{d}_{12})_{LR}|\gg
 |(\delta^{d}_{12})_{RL}|)$} \\ 
 \hline 
0.3& $4.1\times 10^{-4}$ 
 
& $3.0\times 10^{-4}$ 
 
& $3.8\times 10^{-4}$ 
 
& $3.4\times 10^{-4}$ 
\\ 
1.0& $4.6\times 10^{-4}$ 
 
& $3.3\times 10^{-4}$ 
 
& $4.2\times 10^{-4}$ 
 
& $3.7\times 10^{-4}$ 
\\ 
4.0& $6.5\times 10^{-4}$ 
 
& $4.6\times 10^{-4}$ 
 
& $5.8\times 10^{-4}$ 
 
& $5.2\times 10^{-4}$ 
\\ 
\hline  
 $x$ & \multicolumn{4}{c||}{$\sqrt{|\Im  (\delta^{d}_{12} )_{LR}^{2}|} 
\qquad ((\delta^{d}_{12})_{LR} = 
 (\delta^{d}_{12} )_{RL})$} \\ 
 \hline 
0.3& $7.2\times 10^{-4}$ 
 
& $4.4\times 10^{-4}$ 
 
& $3.0\times 10^{-4}$ 
 
& $2.2\times 10^{-4}$ 
\\ 
1.0& $4.1\times 10^{-4}$ 
 
& $3.5\times 10^{-4}$ 
 
& $7.3\times 10^{-4}$ 
 
& $3.7\times 10^{-3}$ 
\\ 
4.0& $4.9\times 10^{-4}$ 
 
& $3.7\times 10^{-4}$ 
 
& $5.0\times 10^{-4}$ 
 
& $4.7\times 10^{-4}$ 
\\ 
\hline  
 $x$ & \multicolumn{4}{c||}{$\sqrt{|\Im  (\delta^{d}_{12} )_{LL}
(\delta^{d}_{12})_{RR}|}$}  \\ 
 \hline 
0.3& $2.3\times 10^{-4}$ 
 
& $1.4\times 10^{-4}$ 
 
& $1.4\times 10^{-4}$ 
 
& $1.1\times 10^{-4}$ 
\\ 
1.0& $2.6\times 10^{-4}$ 
 
& $1.5\times 10^{-4}$ 
 
& $1.5\times 10^{-4}$ 
 
& $1.3\times 10^{-4}$ 
\\ 
4.0& $3.7\times 10^{-4}$ 
 
& $2.1\times 10^{-4}$ 
 
& $2.2\times 10^{-4}$ 
 
& $1.8\times 10^{-4}$ 
\\ 
\hline  \hline 
 \end{tabular} 
 \caption{Limits on $\mbox{Im}\left(\delta_{ij}
 \right)_{AB}\left(\delta_{ij}\right)_{CD}$, with $A,B,C,D=(L,R)$, 
 for an average squark mass $m_{\tilde{q}}=500$ GeV and for different values 
 of $x=m_{\tilde{g}}^2/m_{\tilde{q}}^2$.} 
 \label{tab:imds2_500} 
}

 \TABLE{ 
 \begin{tabular}{||c|c|c|c|c||}  \hline \hline 
 & NO QCD, VIA & LO, VIA & LO, Lattice $B_i$ & NLO, Lattice $B_i$  \\ 
 \hline 
 $x$ & \multicolumn{4}{c||}{$\sqrt{|\Re  (\delta^{d}_{12})_{LL}^{2}|} $} \\ 
\hline 
 0.3& $3.0\times 10^{-2}$ 

& $3.5\times 10^{-2}$ 

& $4.7\times 10^{-2}$ 

& $4.7\times 10^{-2}$ 
\\ 
1.0& $6.4\times 10^{-2}$ 

& $7.4\times 10^{-2}$ 

& $1.0\times 10^{-1}$ 

& $1.0\times 10^{-1}$ 
\\ 
4.0& $1.5\times 10^{-1}$ 

& $1.7\times 10^{-1}$ 

& $2.4\times 10^{-1}$ 

& $2.4\times 10^{-1}$ 
\\ 
\hline  
 $x$ & \multicolumn{4}{c||}{$\sqrt{|\Re  (\delta^{d}_{12} )_{LR}^{2}|} 
 \qquad (|(\delta^{d}_{12})_{LR}|\gg 
|(\delta^{d}_{12})_{RL}|)$} \\ 
\hline 
 0.3& $6.7\times 10^{-3}$ 
 
& $4.7\times 10^{-3}$ 
 
& $6.0\times 10^{-3}$ 
 
& $5.3\times 10^{-3}$ 
\\ 
1.0& $7.4\times 10^{-3}$ 
 
& $5.2\times 10^{-3}$ 
 
& $6.6\times 10^{-3}$ 
 
& $5.9\times 10^{-3}$ 
\\ 
4.0& $1.0\times 10^{-2}$ 
 
& $7.2\times 10^{-3}$ 
 
& $9.2\times 10^{-3}$ 
 
& $8.2\times 10^{-3}$ 
\\ 
\hline  
 $x$ & \multicolumn{4}{c||}{$\sqrt{|\Re  (\delta^{d}_{12} )_{LR}^{2}|} 
\qquad ((\delta^{d}_{12})_{LR} = 
 (\delta^{d}_{12} )_{RL})$} \\ 
 \hline 
 0.3& $1.2\times 10^{-2}$ 
 
& $6.4\times 10^{-3}$ 
 
& $4.5\times 10^{-3}$ 
 
& $3.4\times 10^{-3}$ 
\\ 
1.0& $6.7\times 10^{-3}$ 
 
& $5.7\times 10^{-3}$ 
 
& $1.3\times 10^{-2}$ 
 
& $2.2\times 10^{-2}$ 
\\ 
4.0& $8.0\times 10^{-3}$ 
 
& $5.8\times 10^{-3}$ 
 
& $8.0\times 10^{-3}$ 
 
& $7.5\times 10^{-3}$ 
\\ 
\hline  
 $x$ & \multicolumn{4}{c||}{$\sqrt{|\Re  (\delta^{d}_{12} )_{LL}
(\delta^{d}_{12})_{RR}|} $} \\ 
 \hline 
0.3& $3.8\times 10^{-3}$ 
 
& $2.1\times 10^{-3}$ 
 
& $2.1\times 10^{-3}$ 
 
& $1.8\times 10^{-3}$ 
\\ 
1.0& $4.3\times 10^{-3}$ 
 
& $2.4\times 10^{-3}$ 
 
& $2.4\times 10^{-3}$ 
 
& $2.0\times 10^{-3}$ 
\\ 
4.0& $6.0\times 10^{-3}$ 
 
& $3.3\times 10^{-3}$ 
 
& $3.4\times 10^{-3}$ 
 
& $2.8\times 10^{-3}$ 
\\ 
\hline  \hline 
 \end{tabular} 
 \caption{Limits on $\mbox{Re}\left(\delta_{ij} 
 \right)_{AB}\left(\delta_{ij}\right)_{CD}$, with $A,B,C,D=(L,R)$, for an
  average squark mass $m_{\tilde{q}}=1000$ GeV and for different values
 of $x=m_{\tilde{g}}^2/m_{\tilde{q}}^2$.} 
 \label{tab:reds2_1000} 
} 
 \TABLE{ 
 \begin{tabular}{||c|c|c|c|c||}  \hline \hline 
 & NO QCD, VIA & LO, VIA & LO, Lattice $B_i$ & NLO, Lattice $B_i$  \\ 
 \hline 
 $x$ & \multicolumn{4}{c||}{$\sqrt{|\Im  (\delta^{d}_{12} )_{LL}^{2}|} $} \\ 
\hline 
0.3& $4.0\times 10^{-3}$ 
 
& $4.6\times 10^{-3}$ 
 
& $6.2\times 10^{-3}$ 
 
& $6.2\times 10^{-3}$ 
\\ 
1.0& $8.4\times 10^{-3}$ 
 
& $9.7\times 10^{-3}$ 
 
& $1.3\times 10^{-2}$ 
 
& $1.3\times 10^{-2}$ 
\\ 
4.0& $2.0\times 10^{-2}$ 
 
& $2.3\times 10^{-2}$ 
 
& $3.1\times 10^{-2}$ 
 
& $3.1\times 10^{-2}$ 
\\ 
\hline  
 $x$ & \multicolumn{4}{c||}{$\sqrt{|\Im  (\delta^{d}_{12} )_{LR}^{2}|} 
\qquad (|(\delta^{d}_{12})_{LR}|\gg
 |(\delta^{d}_{12})_{RL}|)$} \\ 
 \hline 
0.3& $8.8\times 10^{-4}$ 
 
& $6.2\times 10^{-4}$ 
 
& $7.9\times 10^{-4}$ 
 
& $7.1\times 10^{-4}$ 
\\ 
1.0& $9.8\times 10^{-4}$ 
 
& $6.9\times 10^{-4}$ 
 
& $8.7\times 10^{-4}$ 
 
& $7.8\times 10^{-4}$ 
\\ 
4.0& $1.4\times 10^{-3}$ 
 
& $9.6\times 10^{-4}$ 
 
& $1.2\times 10^{-3}$ 
 
& $1.1\times 10^{-3}$ 
\\ 
\hline  
 $x$ & \multicolumn{4}{c||}{$\sqrt{|\Im  (\delta^{d}_{12} )_{LR}^{2}|} 
\qquad ((\delta^{d}_{12})_{LR} = 
 (\delta^{d}_{12} )_{RL})$} \\ 
 \hline 
0.3& $1.6\times 10^{-3}$ 
 
& $8.5\times 10^{-4}$ 
 
& $5.9\times 10^{-4}$ 
 
& $4.5\times 10^{-4}$ 
\\ 
1.0& $8.8\times 10^{-4}$ 
 
& $7.6\times 10^{-4}$ 
 
& $1.7\times 10^{-3}$ 
 
& $2.9\times 10^{-3}$ 
\\ 
4.0& $1.1\times 10^{-3}$ 
 
& $7.7\times 10^{-4}$ 
 
& $1.1\times 10^{-3}$ 
 
& $9.9\times 10^{-4}$ 
\\ 
\hline  
 $x$ & \multicolumn{4}{c||}{$\sqrt{|\Im  (\delta^{d}_{12} )_{LL}
(\delta^{d}_{12})_{RR}|}$}  \\ 
 \hline 
0.3& $5.0\times 10^{-4}$ 
 
& $2.8\times 10^{-4}$ 
 
& $2.8\times 10^{-4}$ 
 
& $2.3\times 10^{-4}$ 
\\ 
1.0& $5.6\times 10^{-4}$ 
 
& $3.1\times 10^{-4}$ 
 
& $3.2\times 10^{-4}$ 
 
& $2.6\times 10^{-4}$ 
\\ 
4.0& $8.0\times 10^{-4}$ 
 
& $4.4\times 10^{-4}$ 
 
& $4.5\times 10^{-4}$ 
 
& $3.7\times 10^{-4}$ 
\\ 
\hline  \hline 
 \end{tabular} 
 \caption{Limits on $\mbox{Im}\left(\delta_{ij}
 \right)_{AB}\left(\delta_{ij}\right)_{CD}$, with $A,B,C,D=(L,R)$, 
 for an average squark mass $m_{\tilde{q}}=1000$ GeV and for different values 
 of $x=m_{\tilde{g}}^2/m_{\tilde{q}}^2$.} 
 \label{tab:imds2_1000} 
}

\TABLE{ 
 \begin{tabular}{||c|c|c|c|c||}  \hline \hline 
 & NO QCD, VIA & LO, VIA & LO, Lattice $B_i$ & NLO,  Lattice $B_i$  \\ 
 \hline 
 $m_{\tilde \nu}$ (GeV) & \multicolumn{4}{c||}{$ \Re
 \lambda^\prime_{i21}\lambda^{\prime^*}_{i12} $} \\  
\hline 
 100& $1.4\times 10^{-10}$ 

& $6.1\times 10^{-11}$ 

& $6.3\times 10^{-11}$ 

& $4.4\times 10^{-11}$ 
\\ 
200& $5.8\times 10^{-10}$ 

& $2.2\times 10^{-10}$ 

& $2.3\times 10^{-10}$ 

& $1.6\times 10^{-10}$ 
\\ 
500& $3.6\times 10^{-9}$ 

& $1.2\times 10^{-9}$ 

& $1.3\times 10^{-9}$ 

& $8.5\times 10^{-10}$ 
\\ 
 \hline 

 $m_{\tilde \ell}$ (GeV) & \multicolumn{4}{c||}{$ \Re
 \lambda^\prime_{i31}\lambda^{\prime^*}_{i32} $} \\  

\hline 

200& $6.2\times 10^{-4}$ 

& $2.4\times 10^{-4}$ 

& $2.5\times 10^{-4}$ 

& $1.7\times 10^{-4}$ 

\\ \hline  

 $m_{\tilde \nu}$ (GeV) & \multicolumn{4}{c||}{$ \Im
   \lambda^\prime_{i21}\lambda^{\prime^*}_{i12} $} \\  
\hline 
 100& $3.1\times 10^{-12}$ 

& $1.3\times 10^{-12}$ 

& $1.4\times 10^{-12}$ 

& $9.6\times 10^{-13}$ 
\\ 
200& $1.2\times 10^{-11}$ 

& $4.8\times 10^{-12}$ 

& $4.9\times 10^{-12}$ 

& $3.4\times 10^{-12}$ 
\\ 
500& $7.8\times 10^{-11}$ 

& $2.6\times 10^{-11}$ 

& $2.7\times 10^{-11}$ 

& $1.8\times 10^{-11}$ 
\\ 
\hline  
 $m_{\tilde \ell}$ (GeV) & \multicolumn{4}{c||}{$ \Im
   \lambda^\prime_{i31} \lambda^{\prime^*}_{i32} $} \\ 
\hline 
200& $1.3\times 10^{-5}$ 

& $5.2\times 10^{-6}$ 

& $5.4\times 10^{-6}$ 

& $3.7\times 10^{-6}$ 
\\ 
\hline  \hline 
 \end{tabular} 
 \caption{Limits on $\lambda^\prime_{i21} \lambda^{\prime^*}_{i12}$
   and on $\lambda^\prime_{i31} \lambda^{\prime^*}_{i32}$ from $\Delta
   M_K$ and $\varepsilon_K$ for different values of the relevant SUSY
   masses.}
 \label{tab:rpar} 
} 

\end{document}